\pdfoutput=1
\documentclass[11pt]{article}

\usepackage{acl}
\usepackage{graphicx}
\usepackage{times}
\usepackage{amsmath}
\usepackage{latexsym}
\usepackage{multirow}
\usepackage{placeins}
\usepackage{eufrak}
\usepackage{float}
\usepackage{booktabs}
\usepackage{array}
\usepackage[symbol]{footmisc}

\newcolumntype{L}[1]{>{\raggedright\let\newline\\\arraybackslash\hspace{0pt}}m{#1}}
\newcolumntype{C}[1]{>{\centering\let\newline\\\arraybackslash\hspace{0pt}}m{#1}}
\newcolumntype{R}[1]{>{\raggedleft\let\newline\\\arraybackslash\hspace{0pt}}m{#1}}
\usepackage[T1]{fontenc}
\usepackage[utf8]{inputenc}

\usepackage{microtype}

\title{A Word is Worth A Thousand Dollars: \\ Adversarial Attack on Tweets Fools Stock Predictions
}

\author{{Yong Xie}$^1$, {\bf Dakuo Wang}$^{2, \dagger}$, {\bf Pin-Yu Chen}$^2$, {\bf Jinjun Xiong}$^3$, {\bf Sijia Liu}$^{2,4}$ \and {\bf Sanmi Koyejo}$^1$ \\
$^1${University of Illinois Urbana-Champaign}, $^2${IBM} \\
$^3${University at Buffalo}, $^4${Michigan State University}}

\begin{document}
\maketitle
\begin{abstract}
More and more investors and machine learning models rely on social media (e.g., Twitter and Reddit) to gather real-time information and sentiment to predict stock price movements. Although text-based models are known to be vulnerable to adversarial attacks, whether stock prediction models have similar vulnerability 
is underexplored.
In this paper, we experiment with a variety of adversarial attack configurations to fool three stock prediction victim models.
We address the task of adversarial generation by solving combinatorial optimization problems with semantics and budget constraints. Our results show that the proposed attack method can \textbf{achieve consistent success rates}  and cause \textbf{significant monetary loss} in trading simulation by simply concatenating a perturbed but semantically similar tweet. 

\end{abstract}

\let\thefootnote\relax\footnotetext{\noindent$^{\dagger}$ Corresponding author \href{mailto:dakuo@acm.org}{dakuo@acm.org}. Our code is available at \url{https://github.com/yonxie/AdvFinTweet}}
% \footnote{}

\section{Introduction}
\vspace{-5pt}
% Natural Language Processing (NLP) has become increasingly powerful due to availability of massive amount of data and computing power. 
% Among different applications, financial sentiment analysis and prediction are fast growing since its inception, coinciding with rapid accumulation of social media data on the Internet. Before the surge of social media, the prevailing financial text mining usually involved news articles \cite{schumaker2009textual}, and statistical learning models \cite{fung2003stock}. Social media such as Twitte, facebook, Yahoo Finance Message Board, Stocktwits, etc., soon surge both in terms of users and contents, providing fuels for next stage of natural language analysis. 
The advance of deep learning based language models are playing a more and more important role in the financial context, including convolutional neutral network (CNN) \cite{ding2015deep}, recurrent neutral network (RNN) \cite{minh2018deep}, long short-term memory network (LSTM) \cite{hiew2019bert, sawhney2021fast, hochreiter1997long}, graph neutral network (GNN) \cite{sawhney2020deep, sawhney2020voltage}, transformer \cite{yang2020html}, autoencoder \cite{xu2018stock}, etc. 
% Public sentiment analysis and financial prediction from social media have been fast growing since its inception, coinciding with rapid accumulation of social media data on the internet over the years. 
For example, \citet{antweiler2004all} find that comments on Yahoo Finance can predict stock market volatility after controlling the effect of news. 
\citet{cookson2020don} also show that sentiment disagreement on Stocktwits is highly related to certain market activities. 
% In the meantime, more sophisticated NLP techniques are proposed to appreciate recent advance in deep learning. Deep learning based sentiment analysis and financial forecast is now a core component in this domain. The model zoo includes  
Readers can refer to these survey papers for more details \cite{dang2020sentiment,zhang2018deep,xing2018natural}.
% Therefore, it is reasonable to claim we are standing at the dawn of natural language based financial forecasting. 
\begin{figure}[ht!]
    \centering
    \includegraphics[width=\linewidth]{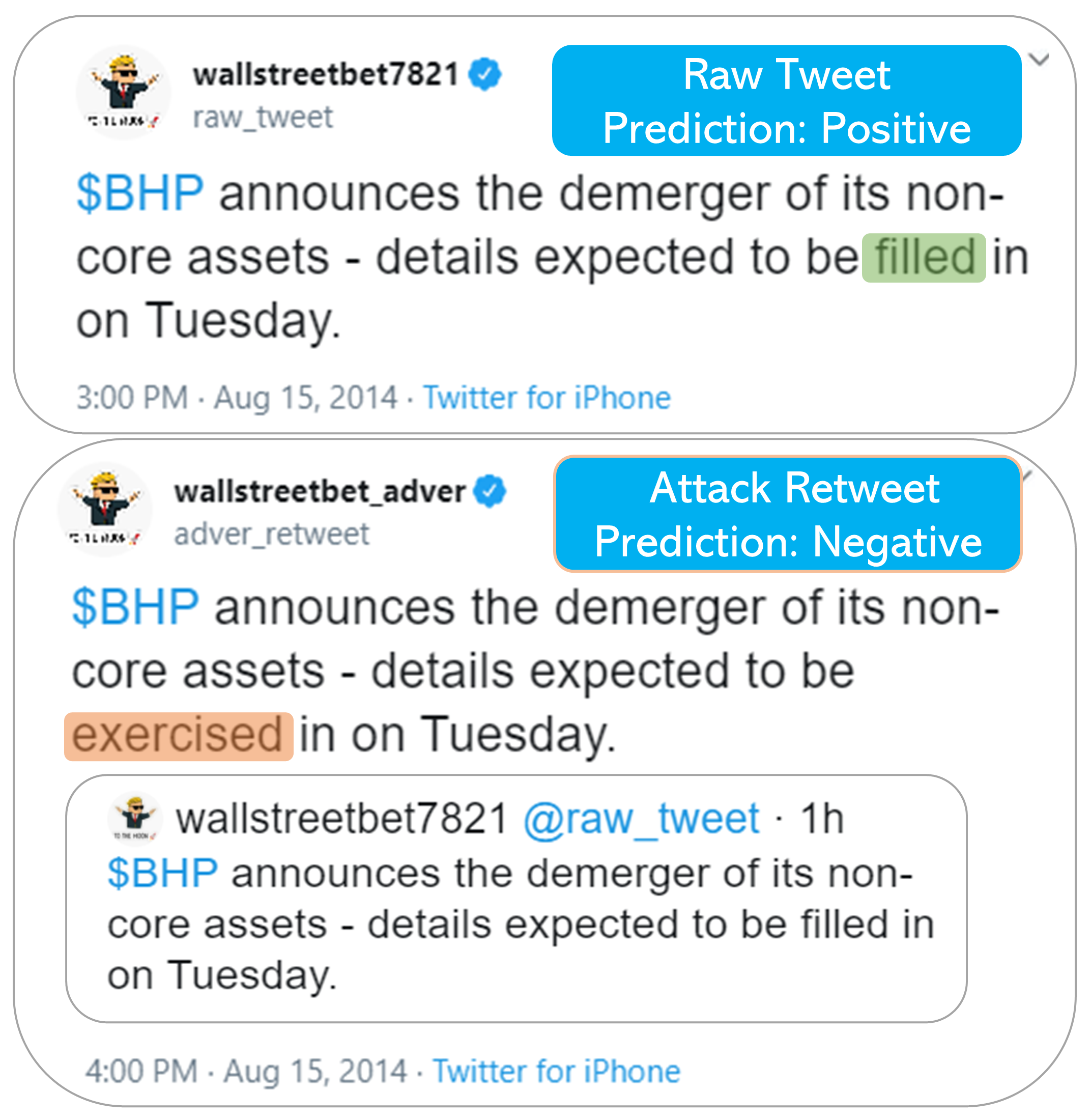}
    \vspace{-22pt}
    \caption{An example of word-replacement adversarial attack. (Top) benign tweet leads \texttt{Stocknet} to predict stock \textcolor{green}{going UP}; (Bottom) adding an adversarial quote tweet leads \texttt{Stocknet} to predict stock \textcolor{red}{going DOWN}.}
    \vspace{-22pt}
    \label{fig:retweet_sample}
% \vspace{12pt}
\end{figure}

It is now known that text-based deep learning models can be vulnerable to adversarial attacks \cite{szegedy2013intriguing, goodfellow2014explaining}. 
The perturbation can be at the sentence level (e.g., \citealp{xu_grey-box_2021, iyyer2018adversarial, ribeiro2018semantically}), the word level (e.g., \citealp{zhang-etal-2019-generating-fluent, alzantot-etal-2018-generating, zang-etal-2020-word, textfoolerjin2020, lei2018discrete, zhang2021crafting, lin_rockner_2021}), or both \cite{chen_multi-granularity_2021}. 
% or the character level (e.g.,  \citealp{belinkov2017synthetic, ebrahimi_hotflip_2018})
% Those models perform comparatively well in terms of both attack success rate and adversarial instance quality, suggesting that perturbation robustness is not guaranteed in NLP model, despite the technical challenges for adversaries. 
We are interested in whether such adversarial attack vulnerability also exists in stock prediction models, as these models embrace more and more human-generated media data (e.g., Twitter, Reddit, Stocktwit, Yahoo News \cite{xu2018stock, sawhney2021fast}).
The adversarial robustness is a more critical issue in the context of stock prediction as anyone can post perturbed tweets or news to influence forecasting models.
For example, a fake news (``Two Explosions in the White House and Barack Obama is Injured'') posted by a hacker using the AssociatedPress's Twitter account on 04/23/2013 erased \$136 billion market value in just 60 seconds \cite{noauthor_syrian_nodate}. Although the event doesn't fall into the category of adversarial attack, it rings the alarm for traders who use (social) media information for their trading decisions. 
% Vulnerability of deep NLP models rings alarm for practitioners, especially for financial analysts and investors who utilize deep NLP models to predict price movements from social media. 
% One the one hand, the data sources for financial forecast model are typically public-traded information, even if data processing and feature engineering can be different. For example, social media data are collected from target platforms, e.g., Twitter and Stocktwit, after applying certain filters , e.g., financial news, containing cashtag, from certain users, etc. 
% The nature of publicity of such platforms amplifies vulnerability of downstream models since adversaries are able to inject adversarial instances by simply uploading maliciously crafted posts without data breaching. 
% As a startling example of potential adversarial risk on social media, 
% official Twitter account of the Associated Press was hacked on April 23, 2013, and sent a fake tweet \textit{Breaking: Two Explosions in the White House and Barack Obama is Injured}, which tipped stock market by \$136 billion in just 1 minute.   

% In this work, we develop a proof-of-concept textual \textbf{F}inancial \textbf{A}dversarial \textbf{A}ttack on \textbf{T}witter, FAAT, by taking physical implementation into consideration. 

To our best knowledge, it is the first paper to consider the adversarial attack in the financial NLP literature. Many attacks modify benign text directly (\textit{manipulation attack}) and use them as model input; However, in our case, adversarial quote tweets enter the model along with benign tweets (\textit{concatenation attack}), which is more realistic as malicious Twitter users can not modify others' tweets. In other words, we formulate the task as a text-concatenating attack \cite{jia-liang-2017-adversarial, le2021sweet}: we implement the attack by injecting new tweets instead of manipulating existing benign tweets. Our task is inspired and mimics the quote tweet function on social media, and uses it to feed the adversarial samples into the dataset. Despite various algorithms are proposed to generate manipulation attack, literature of concatenation attack on classification models is rare, with exceptions \citet{le2021sweet}, \citet{song_universal_2021} and \citet{wang2020t3}. Our paper provides extra evidence of their difference by investigating their performances in the financial domain.

% their transferability to concatenation attack    

The main challenge is to craft new and effective adversarial tweets. 
% While the adversarial tweets can be arbitrary given that they are newly posted, 
We solve the task by aligning the semantics with benign tweets so that the potential human and machine readers can't detect our adversarial tweets. To achieve that, we consider the generation task as a combinatorial optimization problem \cite{zang-etal-2020-word, guo_gradient-based_2021}. Specific tweets are first selected, which are used as the target of perturbation on a limit number of words within the tweets. 
We then examine our attack method on three financial forecast models with \textit{attack success rate}, \textit{F1} and potential \textit{profit and loss} as evaluation metrics. Results show that our attack method consistently achieves good success rate on the victim models. More astonishingly, the attack can cause additional loss of 23\% to 32\% if an investor trades on the predictions of the victim models (Fig.~\ref{fig:pnl}). 
% We conclude the paper with an analysis of the result.

% In summary, our work contributes to: 
% \begin{itemize}
%     \item We propose a concatenation attack that take advantage of re-post feature of social media, formulate it as an optimization task, and develop two solvers to crate adversarial retweets. The attack results demonstrate that financial forecast models are vulnerable to adversarial attack. Our attack can achieve attack success rate of 16\%-30\% with different budgets, and thus leads to sharply drop in model performance and trading profitability.
%     \item  We further examine the difference between concatenation attack and traditional perturbation attack where adversary modifies data instances directly. The results suggest that the two attacks tend to select different tweets and words to apply perturbation. They are not transferable in terms of attack performance and site selection. Therefore, researchers should keep the physical consideration in mind when develop attack models.
%     \item We also analyze the target words selected by our attack model. We find that our attack tends to choose words that fits in the lexicon of financial corpora, instead of perturbing words that carry strong polarization. Although perturbing polarized words may do a better job of fooling human readers, financial machine are easily to be manipulated when perturbation is applied on financial words. 
% \end{itemize}

\section{Adversarial Attack on Stock Prediction Models with Tweet Data}

\paragraph{Attack model: Adversarial tweets.}
% Deep neutral network is more vulnerable when it takes crowd-sourcing data as input. 
% % Traditional adversarial attack requires adversary to directly manipulate dataset, model architecture or parameters, which is usually unrealistic or expensive in the sense of economic cost and legal cost.
% Social media platform enables adversaries to inject adversarial instances by pretending to be data contributors.
In the case of Twitter, adversaries can post malicious tweets which are crafted to manipulate downstream models that take them as input. 
%Ideally, content of adversarial tweet is arbitrary as long as it changes the model prediction. However, irreverent tweets tend to be filtered out in the data pipeline, and fail to attack. 
We propose to attack by posting semantically similar adversarial tweets as quote tweets on Twitter, so that they could be identified as relevant information and collected as model input.
For example, as shown in Fig~\ref{fig:retweet_sample}, the original authentic tweet by the user \textit{wallstreetbet7821} was ``\textit{\$BHP announces the demerger of its non-core assets - details expected to be \textcolor{green}{filled} in on Tuesday.}'' An adversarial sentence could be ``\textit{\$BHP announces the demerger of its non-core assets - details expected to be \textcolor{red}{exercised} in on Tuesday.}''. The outcome of the victim model switches to negative prediction from positive prediction when the quote tweet is added to the input.

The proposed attack method takes the practical implementation into its design consideration, thus has many advantages. First, the adversarial tweets are crafted based on carefully-selected relevant tweets, so they are more likely to pass the models' tweet filter and enter the inference data corpus. Secondly, adversarial tweets are optimized to be semantically similar to the original tweets so that they are not counterfactual and very likely to fool human sanity checks as well as the Twitter's content moderation system. 
% Moreover, retweets with semantically similar content by no means change the landscape of discussion on Twitter, so that our attack remains stealthy at the macro level. 

% \SL{[perturbation, formulation, and methodology can be merged into one section, called 'methodology']}

\paragraph{Attack generation: Hierarchical perturbation.} 

The challenge of our attack method centers around how to select the optimal tweets and the token perturbations with the constraints of semantic similarity. 
In this paper, 
% we settle at word-level attack so that 
we formulate
the task as a \textit{hierarchical perturbation} consisting of three steps: \textit{tweet selection}, \textit{word selection} and \textit{word perturbation}.  In the first step, a set of optimal tweets is first selected as the target tweets to be perturbed and retweeted. 
% The number of tweets are determined by the retweeting budget $b_s$. 
% Traditional attack modifies benign text directly (\textbf{manipulation attack}) and used them as model input; However, in our case, adversarial retweets enter the model along with benign tweets (\textbf{concatenation attack}).  It is more realistic as malicious Twitter users can not modify others' existing tweets, but rather to re-tweet it with a comment.
% Consequently, the selected tweets could be different between the two attack modes.
For each selected tweet in the pool, the word selection problem is then solved to find one or more optimal words to apply perturbation. 
Word and tweet budgets are also introduced to quantify the strength of the perturbation.

We consider the \textbf{word replacemen}t and \textbf{deletion} for word perturbation \cite{garg_bae_2020, li_bert-attack_2020}. In the former case, the final step is to find the optimal candidate as replacement. A synonym as replacement is widely adopted in the word-level attack since it is a natural choice to preserve semantics  \cite{zang-etal-2020-word,dong2021towards, zhang-etal-2019-generating-fluent, textfoolerjin2020}. Therefore, we replace target words with their synonyms chosen from synonym sets which contain the semantically closest words measured by the similarity of the GLOVE embedding \cite{textfoolerjin2020}. 

\paragraph{Mathematical Formulation.} We consider a multimodal stock forecast model $f(\cdot)$ that takes tweet collections $\{\boldsymbol{c_t}\}_{t=1}^T$ and numerical factors $\{\boldsymbol{p_t}\}_{t=1}^T$ as input, where $t$ indexes the date when the data is collected. Peeking into the tweet collection, it contains $|\boldsymbol{c_t}|$ tweets for date $t$, namely, $\boldsymbol{c_t} = \{\boldsymbol{s_t^1},\boldsymbol{s_t^2}, ..., \boldsymbol{s_t^{|{c}_t|}}\}$. Each tweet $\boldsymbol{s_t^i}$ is a text-based sentence of length $|\boldsymbol{s_t^i}|$, denoted as $\boldsymbol{s_t^i} = (w_t^{i,1},...,w_t^{i,j}, ..., w_t^{i, |\boldsymbol{s_t^i}|})$, for $i=1,...,|\boldsymbol{c}_t|$. A directional financial forecast model takes domains of tweets and numerical factors as input, and yields prediction for stocks' directional movement $y\in\{-1,1\}$:
% \vspace{-10pt}
\begin{equation}
    \setlength\abovedisplayskip{5pt}
    \hat{y}_{t+1} = f(\boldsymbol{c_{t-h:t}}, \boldsymbol{p_{t-h:t}}),
    \setlength\belowdisplayskip{5pt}
\end{equation}
% where 
$h$ is the looking-back window for historical data.

The hierarchical perturbation can be cast as a combinatorial problem for tweet selection $\boldsymbol{m}$, word selection $\boldsymbol{z}$ and replacement selection $\boldsymbol{u}$. 
% Let $\boldsymbol{c_t'}$ be the perturbed tweet collection at time $t$ created by solving the hierarchical perturbation problem. 
The boolean vector $\boldsymbol{m}$ indicates the tweets to be selected. 
% If $m_i=1$, then $i$-th tweet is the target tweet to be perturbed and retweeted. 
For $i$-th tweet, vector $\boldsymbol{z_i}$ indicates the word to be perturbed. As for the word perturbation task, another boolean vector $\boldsymbol{u_{i,j}}$ selects the best replacement. 
% $n_m$ and $n_z$ and $n_u$ denote the maximum amount of tweets, maximum amount of words in each tweet, and the amount of synonyms for each word, respectively. 
% We identify deletion perturbation as a special case of replacement with $u_{i,j,k}=1$ only for padding token, so that the task degenerates to tweet selection and word selection. 
It follows that the hierarchical perturbation can be formulated as
\begin{equation} \label{eq:perturbaton}
    \boldsymbol{c_t'} =  (\boldsymbol{1}-\boldsymbol{m}\cdot \boldsymbol{z}) \cdot \boldsymbol{c_t} + \boldsymbol{m} \cdot \boldsymbol{z} \cdot \boldsymbol{u} \cdot {S}(\boldsymbol{c_t}),
\end{equation}
where $\cdot$ denotes element-column wise product, $\boldsymbol{m}\cdot \boldsymbol{z}$ indicates the selected words in selected tweets, $\boldsymbol{m} \cdot \boldsymbol{z} \cdot \boldsymbol{u}$ indicates selected synonyms for each selected word, and ${S}(\cdot)$ is a element-wise synonym generating function.
% \begin{equation} 
%     \setlength\abovedisplayskip{5pt}
%     \begin{split}
%          \boldsymbol{c_t'} & =  (\boldsymbol{1}-\boldsymbol{m}\cdot \boldsymbol{z}) \cdot \boldsymbol{c_t} + \boldsymbol{m} \cdot \boldsymbol{z} \cdot \boldsymbol{u} \cdot \mathbb{S}(\boldsymbol{c_t}) \\
%         s.t. & \quad \boldsymbol{1}^T\boldsymbol{m} \leq b_{s}, \\
%         & \quad \boldsymbol{1}^T\boldsymbol{z_i} \leq b_{w},\forall i, \\
%         & \quad \boldsymbol{1}^T\boldsymbol{u_{i,j}} = 1, \forall i,j,
%     \end{split}
%     \setlength\belowdisplayskip{0pt}
% \end{equation}
Consequently, given attack loss $\mathcal{L}$, generation of adversarial quote tweets can be formulated as the  optimization program $ \underset{\boldsymbol{m,z,u}}{\min} \mathcal{L}(\boldsymbol{c_t'}\cup \boldsymbol{c_{t-h:t}}, \boldsymbol{c_{t-h:t}} | \boldsymbol{p_{t-h:t}}, f)$,
% \begin{align}\label{opt:raw_optimization}
%     \begin{array}{cl}
%          \underset{\boldsymbol{m,z,u}}{\min} & \mathcal{L}(\boldsymbol{c_t'}\cup \boldsymbol{c_{t-h:t}}, \boldsymbol{c_{t-h:t}} | \boldsymbol{p_{t-h:t}}, f)  \\ 
%           s.t. & \text{budget constraints},
%     \end{array}
% \end{align}
subject to the budget constraints: \textit{a)} $\boldsymbol{1}^T\boldsymbol{m} \leq b_{s}$, \textit{b)} $\boldsymbol{1}^T\boldsymbol{z_i} \leq b_{w},\forall i$ and \textit{c)} $\boldsymbol{1}^T\boldsymbol{u_{i,j}} = 1, \forall i,j$, where $b_s$ and $b_w$ denote the tweet and word budgets. It is worth to stress that perturbation is only applied to the date ($t$) when the attack is implemented to preserve the temporal order. 

To solve the program, we follow the convex relaxation approach developed in \cite{srikant2021generating}. Specifically, the boolean variables (for tweet and word selection)
are relaxed into the continuous space so that they can be optimized by gradient-based methods over a convex hull. Two main implementations of the optimization-based attack generation method are proposed: \textit{joint optimization} (JO) solver and \textit{alternating greedy optimization}   (AGO) solver. JO calls projected gradient descent method to optimize the 
tweet and word selection variables and word replacement variables simultaneously. AGO uses an alternative optimization procedure to sequentially update the discrete selection variables and the replacement selection variables. More details on the optimization program and the solvers can be found in Appendix \ref{appendix:math_formation}.

\section{Experiments}
\vspace{-5pt}

\paragraph{Dataset \& victim models.} We evaluate our adversarial attack on a stock prediction dataset consisting of 10,824 instances including relevant tweets and numerical features of 88 stocks from 2014 to 2016 \cite{xu2018stock}. 
Three models (\textbf{\texttt{Stocknet}} \cite{xu2018stock}, \textbf{\texttt{FinGRU}} based on GRU \cite{cho_learning_2014} and \textbf{\texttt{FinLSTM}} based on LSTM \cite{hochreiter1997long}) of binary classification are considered as victims in this paper.
We apply our attack to instances on which the victim models make correct predictions.

\paragraph{Evaluation metrics.} Attack performance is evaluated by two metrics: \textit{Attack Success Rate} (ASR) and victim model's \textit{F1} drop after attack. ASR is defined as the percentage of the attack efforts that changes the model output. The two metrics gauge the efficacy of the attack and its impact on model performance:
%ASR characterizes the capability of the attack model, and higher the ASR, better the attack. 
% F1 indicates the prediction performance of the victim model, and the pre-attack F1 is 1. The drop of the F1 score of a model demonstrates the success of the attack method. 
% The difference between F1-score before and after deploying our attack demonstrates the direct impact on the prediction performance. 
More efficient attack leads to higher ASR and more decline of F1. 
%Since we only consider the samples that are correctly predicted, The F1-score in the case of no attack is 1.
Moreover, we simulate a \textbf{\textit{Long-Only Buy-Hold-Sell}} strategy \cite{sawhney2021fast, feng2019temporal} with victim models, and calculate the {\textit{Profit and Loss}} (PnL) for each simulation. 
% as the metric the gauge the monetary impact of our attack. 
% This widely-used financial indicator measures the profitability of a trading strategy.
% There are many trading strategies can be used together with a binary classification model, and in our paper, we use the simple 
% This trading strategy \textit{buy} stock(s) on Day $T$ if the model predicts these stocks go up on Day $T+1$, \textit{hold} for one day, and \textit{sell} these stocks the next day no matter what prices will be, and repeat it. 
% It does not \textit{short} a stock even when the model predict a negative move in the second day.  
% Consequently, traders monetize the financial forecast model by betting on the prediction for stock movement at $t+1$. 
Assume a portfolio starts with initial net value \$10000 (100\%), its net value at the end of test period reflects the profitability of the trading strategy and the underlying model. Consequently, the change in PnLs measures the monetary impact of our attack.  
% By including the PnL as evaluation metric, we are able to glance into the impact of our attack on the level of trading strategy and investment portfolio.  
More details on the dataset, victim models and evaluation metrics are housed in Appendix \ref{appendix:experiment_setting}.
\section{Results}
\vspace{-5pt}
\paragraph{Attack performance with single perturbation. } The experiment results for the concatenation attack with word replacement perturbation is shown in Table \ref{tab:overall_results} (with tweet and word budgets both as 1).
% he t, that is, the adversary can only posts 1 retweet which has only 1 word replacement in it. 
%NA stands for no attack, so that the ASR is 0 and F1-score is 1. RA stands for random attack, namely random tweet and word are selected to apply perturbation. Random attack characterizes the model's sensitivity to the tweet collecting procedure; non-zero ASR of random attack implies that inclusion and exclusion of tweets may change the model outcome.
 % JO and AGO correspond to our proposed attack generation methods. 
% We run JO and AGO for 10 iterations to ensure fair comparison. Attack performance is boosted when the site selection are optimized by either JO or AGO. 
% As we can see, 
For both JO and AGO, ASR increases by roughly 10\% and F1 drops by 0.1 on average in comparison to the random attack. 
Such performance drop is considered significant in the context of stock prediction given that the state-of-the-art prediction accuracy of interday return is only about 60\%.

\begin{table}[htb]
\centering
% \begin{tabular}{lC{1.3cm}C{1.3cm}C{1.3cm}C{1.3cm}|C{1.3cm}C{1.3cm}C{1.3cm}C{1.3cm}}
\resizebox{0.95\linewidth}{!}{
\begin{tabular}{lC{0.8cm}C{0.8cm}C{0.8cm}C{0.9cm}|C{0.8cm}C{0.8cm}C{0.8cm}C{0.9cm}}
\toprule
\multicolumn{1}{c}{\multirow{2}{*}{\textbf{Model}}} & \multicolumn{4}{c|}{\textbf{ASR(\%)}} & \multicolumn{4}{c}{\textbf{F1}} \\ \cmidrule(l){2-9} 
\multicolumn{1}{c}{}                                & NA   & RA   & JO   & AGO  & NA  & RA  & JO & AGO \\ 
\hline
Stocknet & 0 & 4.5  & \textbf{16.8} & 11.8 & 1 & 0.96 & \textbf{0.84} & 0.88 \\
FinGRU   & 0 & 5.1  & \textbf{16.4} & 14.1 & 1 & 0.95 & \textbf{0.85} & 0.87 \\
FinLSTM  & 0 & 11.9 & 16.5 & \textbf{19.7} & 1 & 0.89 & 0.85 & \textbf{0.78} \\
% HAN      & 0 & 3.3  & 3.2  & 3.9     & 1 & 0.97 & 0.97 &      \\ 
\bottomrule
\end{tabular}
}
\caption{{Performance of the various adversarial attacks.} NA: no attack; RA: random attack; JO: joint optimization; and AGO: alternating greedy optimization.}
\label{tab:overall_results}
\vspace{-15pt}
\end{table}

\paragraph{Effect of attack budget. }

\begin{figure}[h]
    \vspace{-10pt}
    \centering
    \includegraphics[width=0.49\textwidth]{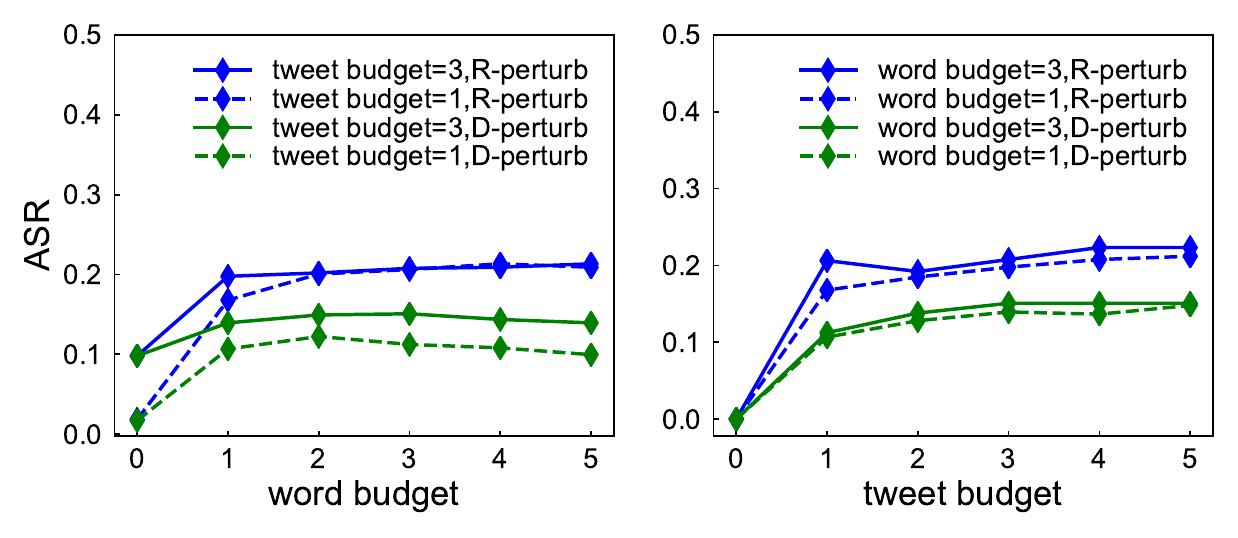}
    \vspace{-23pt}
    \caption{{Effect of attack budgets on ASR with Stocknet as victim model and with JO solver}. r-perturb: word replacement; d-perturb: word deletion. }
    \label{fig:budget_effect}
    \vspace{-15pt}
\end{figure}

We report the effect of different attack budgets on the attack performance in Fig. \ref{fig:budget_effect}. 
We observe that the more budgets allowed (perturbing more tweets and words), the better the attack performance, but the increase is not significant. 
% As shown in Figure \ref{fig:budget_effect}, we also observe the same pattern in our experiment. 
% \SL{[Update the following, do not use math.]}
% The left panel demonstrates the effect of word budget $b_{z}$ in the case of $b_{m}=3$ and $b_{m}=1$. It is worth to stress that we simply replicate the selected tweet(s) if word budget is 0. Analogically, the right panel focuses on effect of tweet budget for $b_{z}=3$ and $b_{z}=1$.
It appears that the attack performance becomes saturated if we keep increasing the attack budgets. In fact, the attack with budget of one tweet and one word is the most cost effective, provided that it introduces minimum perturbation but achieves a relatively similar ASR. 
% By comparing attack performance of two types of word perturbation: replacement (R) and deletion (D),  Replacement perturbation adds another layer of flexibility by choosing different synonyms, so that it leads to better performance in our experiment. However, the pattern of effect of attack budget holds for deletion perturbation as well. 
%However, we only observe slight improvement on attack performance, if not none, even if we extend  budget by five times. It implies that our attack is budget-insensitive. In other words, our attack could achieve  similar attack performance even if it flies on small budget. 

\paragraph{Manipulation vs concatenation attack. }

\begin{figure}[h]
    \vspace{-10pt}
    \centering
    \includegraphics[width=0.49\textwidth]{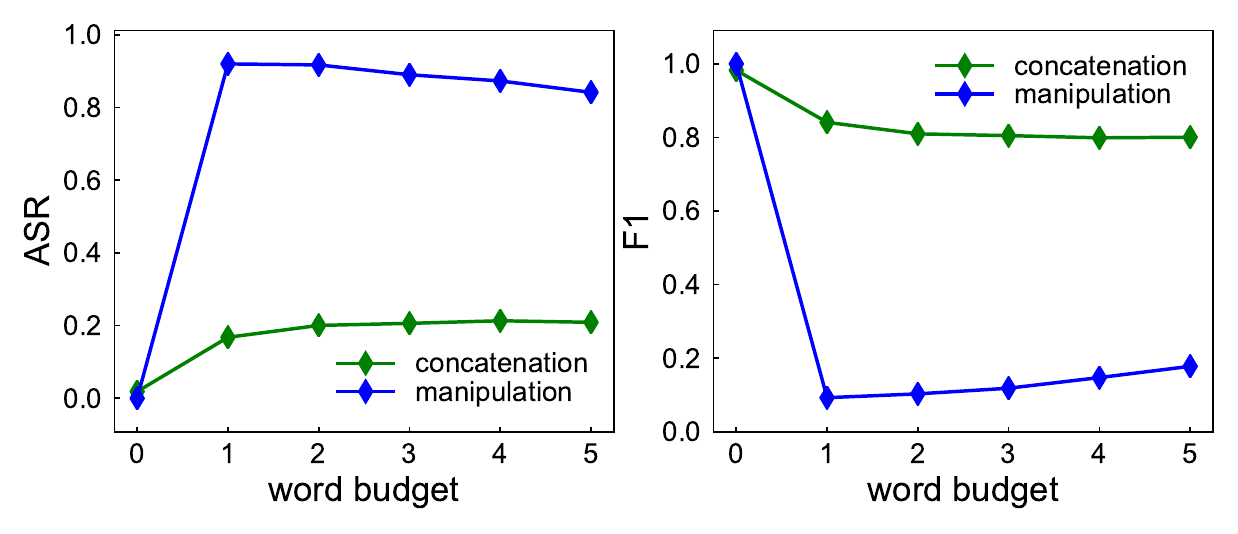}
    \vspace{-23pt}
    \caption{Comparison between \textbf{manipulation} and \textbf{concatenation} attacks with \textbf{word-replacement} perturbation method. \texttt{Stocknet} is the victim model.}
    \label{fig:concatenation_manipulation}
    \vspace{-10pt}
\end{figure}

We focus on concatenation attack in this paper since we believe it is distinct from manipulation attack. We investigate the difference by applying the same method of tweet generation to implement manipulation attack, where the adversarial tweets replace target tweets instead. The experiment runs with one word budget and one twee budget, and the results are reported in Fig. \ref{fig:concatenation_manipulation}. 

It is clear that manipulation attack remarkably outperforms concatenation attack in terms of ASR and F1. Even though the success rate of concatenation attack lags behind the state-of-the-art textual attack, the manipulation attack achieves performance of the same ballpark, which demonstrates the efficacy of optimization-based attack and our solvers. More importantly, it implies that the attack is not transferable between the two tasks, documenting more evidence on language attack transferability \cite{yuan2021transferability, he_model_2021}. The bottom line is that they are two different tasks under different assumptions. Researchers should take downstream scenarios into account when develop attack models. 

\begin{figure}[hb!]
    \centering
    \vspace{-10pt}
    \includegraphics[width=0.45\textwidth]{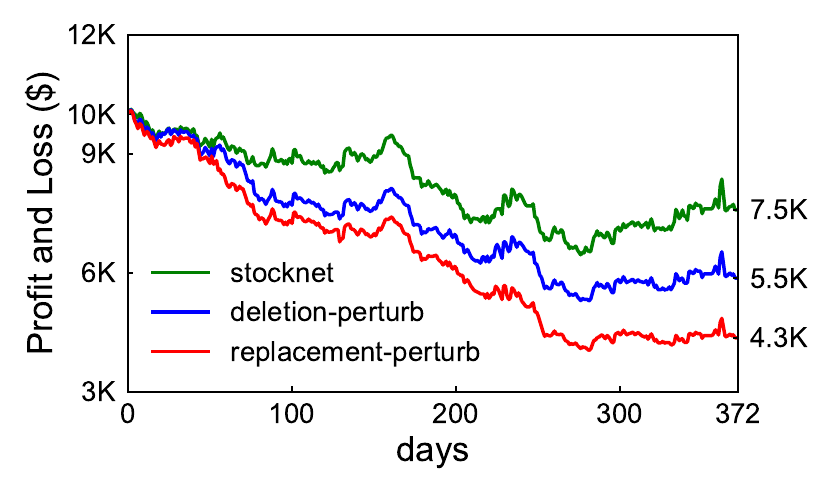}
    \vspace{-15pt}
    \caption{{\textit{Profit and Loss} with \texttt{Stocknet} as the victim model using \textit{Long-Only Buy-Hold-Sell }strategy for 2 years} with \$10K initial investment. \textcolor{green}{Green line}: trading using \texttt{Stocknet} without attack; \textcolor{blue}{Blue line}: concatenation attack with \textbf{deletion perturbation}; \textcolor{red}{Red line}: concatenation attack with \textbf{replacement perturbation}. }
    \vspace{-15pt}
    \label{fig:pnl}
\end{figure}

\paragraph{Trading simulation.} 
% The impact of adversarial attack doesn't end with wrong prediction, but extend to downstream financial analysis and tradings. 
The ultimate measure of a stock prediction model's performance is profitability.
% We adopt a simple long only buy-hold-sell strategy to showcase the the potential damage that can be done through adversarial attack. 
Fig.~\ref{fig:pnl} plots the \textit{profit and loss} of the same trading strategy with \texttt{Stocknet} as the prediction model with or without  the attack -- JO is used to generate adversarial quote tweets. 
% The results on the other two victim models are listed in Appendix. 
For each simulation, the investor has \$10K (100\%) to invest;
% at the beginning
% It is clear that PnL drops sharply if the model is under attack; PnL based on FinGRU tumbles most by 29.7\% from 0.96 to 0.63, and PnL based on FinLSTM falls least by 14.2\% from 0.89 to 0.66. 
% The results may be exaggerated since we run the simulation under ideal assumption, however, even tiny drop in portfolio PnL can leads to loss of millions of dollars, if not more. 
the results show that the proposed attack method with a quote tweet with only a single word replacement can cause the investor an additional \$3.2K (75\%-43\%) loss to their portfolio after about 2 years.

% The investors and institutions who rely on text-based stock prediction models should be alerted.

\section{Conclusion}
\vspace{-5pt}
% We show that financial forecast models are vulnerable to adversarial attacks even with physical constraints.
% a stock prediction adversarial attack problem on Twitter, or {FAAT}, and propose two solvers to create adversarial retweets. 
This work demonstrates that our adversarial attack method consistently fools various financial forecast models even with physical constraints that the raw tweet can not be modified. 
Adding a quote tweet with only one word replaced, the attack can cause 32\% additional loss to our simulated investment portfolio. 
% We also analyze the properties of our attack, including effect of attack budget, characteristics of target words and relation to traditional manipulation attack. 
Via studying financial model's vulnerability, \textbf{our goal is to raise financial community's awareness of the AI model's risks}, so that in the future we can develop more robust human-in-the-loop AI architecture~\cite{wang2019human} to cope with this and other real-world attacks, including black-box attack, unknown input domains, etc.   

\clearpage
\newpage
\bibliography{anthology,reference_normalized}
\bibliographystyle{acl_natbib}

\clearpage
\newpage

\appendix

\section{Mathematical Formation}\label{appendix:math_formation}

\subsection{Financial Forecast Model} 

Massive amounts of text data are generated by millions of users on Twitter every day. 
Among a variety of discussion, stock analysis, picking and prediction are consistently one of the trending topics.
And investors often use the Twitter \textit{cashtag} function (a \$ symbol followed by a ticker) to organize their particular thoughts around one single stock, e.g., \$AAPL, 
so that users can click and see the ongoing discussions. Textual data on Twitter is collectively generated by all of its users via posting tweets. 
Financial organizations and institutional investors often ingest the massive text data in real time and incorporate them or their latent representation into their stock prediction models.
% utilizing it in sentiment analysis, financial prediction, etc. 
% However, not all the tweets are relevant to financial forecast, so tweet filters might be applied in the tweet collection pipeline which eliminate irrelevant tweets. 
% Relevant tweets are then processed to generate features along with price data, which are further sent to downstream models for analytics. 

% \begin{figure}[h]
%     \centering
%     \includegraphics[width=0.5\textwidth]{fig/adv_paper.pdf}
%     \vspace{-40pt}
%     \caption{Data pipeline}
%     \label{fig:tweet_pipeline}
% \end{figure}

% We lay out the mathematical formation for the proposed attack model in this section. 
We consider the multimodal stock forecast models that take tweet collections $\{\boldsymbol{c_t}\}_{t=1}^T$ and numerical factors $\{\boldsymbol{p_t}\}_{t=1}^T$ as input,where $t$ indexes the date when the data is collected. The numerical factors are usually mined from historical price, fundamentals and other alternative data sources. In this paper, we assume that the domain of numerical factors is unassailable since they are directly derived from public records. Therefore, the objective of adversary is to manipulate model output by injecting perturbation to the textual domain $\{\boldsymbol{c_t}\}_{t=1}^T$. Peeking into the tweet collection, it contains $|\boldsymbol{c_t}|$ tweets for date $t$, namely, $\boldsymbol{c_t} = \{\boldsymbol{s_t^1},\boldsymbol{s_t^2}, ..., \boldsymbol{s_t^{|{c}_t|}}\}$. Each tweet $\boldsymbol{s_t^i}$ is a text-based sentence of length $|\boldsymbol{s_t^i}|$, denoted as $\boldsymbol{s_t^i} = (w_t^{i,1},...,w_t^{i,j}, ..., w_t^{i, |\boldsymbol{s_t^i}|})$, for $i=1,...,|\boldsymbol{c}_t|$. A directional financial forecast model takes domains of tweets and numerical factors as input, and yields prediction for stocks' directional movement $y\in\{-1,1\}$:
\begin{align}
    \hat{y}_{t+1} = f(\boldsymbol{c_{t-h:t}}, \boldsymbol{p_{t-h:t}}),
\end{align}
where $h$ is the looking-back window for historical data.

\subsection{Attack Model}
Let $\boldsymbol{c_t'}$ be the perturbed tweet collection at time $t$ created by solving the hierarchical perturbation problem. To formalize the perturbation task, we introduce boolean vector variable $\boldsymbol{m}\in \{0,1\}^{n_{m}}$ to indicate the tweets to be selected. If $m_i=1$, then $i$-th tweet is the target tweet to be perturbed and retweeted. Besides, for $i$-th tweet, vector $\boldsymbol{z_i}\in\{0,1\}^{n_{z}}$ indicates the word to be perturbed. As for the word perturbation task, another boolean vector $\boldsymbol{u_{i,j}} \in \{0,1\}^{n_{u}}$ selects the best replacement. $n_m$ and $n_z$ and $n_u$ denote the maximum amount of tweets, maximum amount of words in each tweet, and the amount of synonyms for each word, respectively. We identify deletion perturbation as a special case of replacement with $u_{i,j,k}=1$ only for padding token, so that the task degenerates to tweet selection and word selection. Let vector $\boldsymbol{z} \in \{0,1\}^{n_{m}\times n_{z}}$ denote $n_{m}$ different $\boldsymbol{z_i}$ vector, and $\boldsymbol{u}\in \{0,1\}^{n_{m}\times n_{z}\times n_{u}}$ denote $n_{m}\times n_{z}$ different $u_{i,j}$ vectors. It follows that the hierarchical perturbation can be defined as
\begin{align} \label{appendix_eq:perturbaton}
    \begin{split}
        % {\textstyle
         \boldsymbol{c_t'} & =  (\boldsymbol{1}-\boldsymbol{m}\cdot \boldsymbol{z}) \cdot \boldsymbol{c_t} + \boldsymbol{m} \cdot \boldsymbol{z} \cdot \boldsymbol{u} \cdot {S}(\boldsymbol{c_t}) \\
        s.t. & \quad \boldsymbol{1}^T\boldsymbol{m} \leq b_{s}, \\
        & \quad \boldsymbol{1}^T\boldsymbol{z_i} \leq b_{w},\forall i, \\
        & \quad \boldsymbol{1}^T\boldsymbol{u_{i,j}} = 1, \forall i,j,
        %  } 
    \end{split}
\end{align}
where $\cdot$ denotes element-column wise product, $b_s$ denotes tweet budget, $b_w$ denotes word budget and ${S}(\cdot)$ is element-wise synonym generating function. 

Adversarial quote tweets are then passed into downstream financial forecast model $f(\cdot)$ along with benign tweets. Attack success is achieved if the adversarial tweets manage to fool the downstream model, and change the model output. Financial forecast model usually takes observation of multiple steps as input to appreciate the temporal dependence. However, adversary can only inject adversarial quote tweets at present time. That is, when run the model on day $t$ to predict price movement on day $t+1$, quote tweets only enter tweet collection for day $t$; collections for days prior to $t$ remain static. Consequently, generation of successful adversarial quote tweets is formulated as the following optimization program:
\begin{align}\label{opt:raw_optimization}
    \begin{array}{cl}
         \underset{\boldsymbol{m,z,u}}{\min} & \mathcal{L}(\boldsymbol{c_t'}\cup \boldsymbol{c_{t-h:t}}, \boldsymbol{c_{t-h:t}} | \boldsymbol{p_{t-h:t}}, f)  \\ 
          s.t. & \text{constraint in } \eqref{appendix_eq:perturbaton},
    \end{array}
\end{align}
where $\mathcal{L}$ denotes the attack loss. We adopt the cross-entropy loss for our attack since it is untargeted attack \cite{srikant2021generating}. Other classification-related loss may be applied according to adversary's objective. Furthermore, we also add entropy-based regularization to encourage sparsity of optimization variables \cite{dong2021towards}. 
\subsection{Methodology}
The challenge of solving program \eqref{opt:raw_optimization} lies in the combinatorial and hierarchical nature. We first relax the boolean variables into continuous space so that they can be solved by gradient-based solvers. A common workaround for combinatorial optimization is to solve an associated continuous optimization over convex hull \cite{dong2021towards, srikant2021generating}. An computationally efficient fashion is to optimize over a convex hull constructed with linear combination of candidate set, and the optimal replacement goes with word with highest weight \cite{dong2021towards}. However, this approach doesn't fit in the hierarchical tweet and word selection problem. For example, in order to select the optimal target word, one need to sum over the embedding of all words in the tweet, so the tweet collapses into embedding for one hypothetical word. Similarly, different tweets collapse to one hypothetical tweet, or one hypothetical word when one jointly selects tweets and words. 

\paragraph{Joint optimization solver (JO).} As a remedy, we propose a \textit{joint optimization solver} that combines projected gradient descent and convex hull to jointly optimize $\boldsymbol{m}$, $\boldsymbol{z}$ and $\boldsymbol{u}$. Replacement selection is optimized over the convex hull:
\begin{align*}
    {\textstyle
    \boldsymbol{c'_t} = (1-\boldsymbol{m}\cdot\boldsymbol{z})\cdot\boldsymbol{c_t} + \boldsymbol{m}\cdot\boldsymbol{z} \cdot conv(\boldsymbol{u}, S(\boldsymbol{\boldsymbol{c_t}})),}
\end{align*}
where 
\begin{align*}
    conv(\boldsymbol{u},S(\boldsymbol{\boldsymbol{c_t}}))=\{\sum_k\hat{u}_{i,j,k}{S}(w_{i,j,k}), \forall i, j\}, 
\end{align*}
and 
\begin{align*}
    \hat{u}_{i,j,k} = \frac{exp(u_{i,j,k})}{\sum_k exp(u_{i,j,k})}.
\end{align*}
The problem of \eqref{opt:raw_optimization} is then solved by optimizing $\hat{\boldsymbol{u}}$. Unlike $\boldsymbol{u}$, $\boldsymbol{m}$ and $\boldsymbol{z}$ are optimized directly via projected gradient descent (PGD).  Moreover, when $\boldsymbol{m}$ is one-hot vector, it determines the tweets to be retweeted, and those quote tweets are then added into tweet collection. However, $\boldsymbol{m}$ is continuous during optimization,  so we retweet all the collected tweets and add them into tweet collection, which helps generate and back-propagate gradients for all the entries of $\boldsymbol{m}$. After the optimization is solved, we map the continuous solution into one-hot vector by selecting top $b_{s}$ highest $m_i$.

\paragraph{Alternating greedy optimization solver (AGO).} 
Greedy optimization is usually computational ineffective since a vast amount of inquiries is required when we collect large amount of tweets and have high attack budget. To mitigate the problem, we alternate the optimization over $\boldsymbol{m}$, $\boldsymbol{z}$ and $\boldsymbol{u}$. The aforementioned convex hull approach is adopted for finding optimal $\boldsymbol{u}$. The difference lies on the path to solve tweet and word selection problems. More specifically, we alternatively search the optimal target tweets and words which achieve the highest increases in prediction loss. For tweet selection, we mimic the physical attack scenario, and new quote tweets are added into tweet collection during the greedy search. Depending on the adversary's objective, different metrics may be used to measure the importance of each tweet and word. For example, \citet{alzantot-etal-2018-generating} use predicting probability to determine the selection of words; \citet{ren2019generating} propose probability weighted word saliency as criterion for word selection; \citet{textfoolerjin2020} calculate the prediction change before and after deletion as word importance. 

\section{Experimental Settings} \label{appendix:experiment_setting}
\subsection{Dataset}
We evaluate our adversarial attack on a stock prediction dataset \cite{xu2018stock}. 
The dataset contains both tweets and historical prices (e.g., open, close, high, etc) for 88 stocks of 9 industries: \textit{Basic Materials, Consumer Goods, Healthcare, Services, Utilities, Conglomerates, Financial, Industrial Goods and Technology.} 
% Those stocks are top tickers in terms of capital size since stocks of high volume tend to attract more discussion on Twitter. 
Since we consider the task of binary classification, data instances are supposed to labelled positive and negative for upward and downward movement respectively. 

Moreover, it is observed that the dataset contains a number of instances with exceptionally minor price movements. In practice, minor movement is hard to be monetized due to the existence of transaction cost. Therefore, an upper threshold of 0.55\% and a lower threshold of -0.5\% are introduced. Specifically, stocks going up more than 0.55\% in a day are labeled as positive, those going down more than -0.5\% are labeled as negative, and the minor moves in between are filtered out. As argued in \cite{xu2018stock}, the particular thresholds are carefully selected to balance the two classes. 

In addition, the sampling period spans from 01/01/2014 to 01/01/2016. We split the dataset into train and test set on a rolling basis. This special program improves the similarity between distributions of train set and test set, which is widely adopted on temporal dataset. It leaves us 9416 train instances and 1408 test instances in 7 nonconsecutive periods. For the text domain, the dataset contains 57533 tweets in total. 
%  \vspace{-5pt}
% \begin{table}[!h]
%     \centering
   
%     \resizebox{0.95\linewidth}{!}{
%     \begin{tabular}{c|c|c|c}
%     \toprule
%     \textbf{Period} & \textbf{Dataset} & \textbf{Sample Num.} & \textbf{Tweet Num.} \\
%     \hline
%     2014/01/01-2014/03/31 & train & 1 & 1 \\
%     \hline
%     2014/04/01-2014/04/15 & test & 1 & 1 \\
%     \hline
%     2014/04/16-2014/07/15 & train & 1 & 1 \\
%     \hline
%     2014/07/16-2014/07/31 & test & 1 & 1 \\
%     \hline
%     2014/08/01-2014/10/31 & train & 1 & 1 \\
%     \hline
%     2014/11/01-2014/11/15 & test & 1 & 1 \\
%     \hline
%     2014/11/16-2015/02/15 & train & 1 & 1 \\
%     \hline
%     2015/02/16-2014/02/28 & test & 1 & 1 \\
%     \hline
%     2015/03/01-2015/05/30 & train & 1 & 1 \\
%     \hline
%     2015/06/01-2015/06/15 & test & 1 & 1 \\
%     \hline
%     2015/06/16-2015/09/15 & train & 1 & 1 \\
%     \hline
%     2015/09/16-2015/09/30 & test & 1 & 1 \\
%     \hline
%     2015/10/01-2015/12/15 & train & 1 & 1 \\
%     \hline
%     2015/12/16-2015/12/31 & test & 1 & 1 \\
%     \hline
%     \end{tabular}
%     }
%     \caption{Dataset partition and statistics.}
%     \label{tab:my_label}
% \end{table}

\subsection{Victim Models}
\paragraph{Stocknet.} A variational Autoencoder (VAE) that takes both tweets and price as input \cite{xu2018stock}. Tweets are encoded in hierarchical manner within days, and then modeled sequentially along with price features. It consists of three main components in bottom-up fashion. Market Information Encoder first encodes tweets and prices to a latent representation of 50 dimensions for each day. Variational Movement Decoder infers latent vectors of 150 dimensions and then decodes stock movements. At last, a module called Attentive Temporal Auxiliary integrates temporal loss through an attention mechanism. We train the model on the dataset from scratch with the same configurations as \citet{xu2018stock}.

\paragraph{FinGRU.} A binary classifier that takes numerical features and tweets as input. All features are encoded sequentially by GRU \cite{cho_learning_2014} to exploit the temporal dependence. The model adopts the same Market Information Encoder as Stocknet. Latent representation of tweets and prices are then fed into a layer of GRU with attention mechanism to integrate temporal information. We train the model with an Adam optimizer \cite{adam} and learning rate of 0.005. The checkpoint achieves the best performance on test dataset among 100 epochs is adopted as the victim model. 

\paragraph{FinLSTM.} A binary classifier identical to FinGRU, but utilizes LSTM \cite{hochreiter1997long} to encode temporal dependence. The model is trained in the same manner as FinGRU. 

\subsection{Evaluation Metrics}
Following \citet{srikant2021generating}, we evaulate the attack on those examples in the test set that are correctly classified by the target models. It provides direct evidence of the adversarial effect of the input perturbation and the model robustness. In the specific application of financial forecast, it makes more sense to manipulate correct prediction than incorrect ones. The following two common metrics are adopted to evaluate attack performance. 
\paragraph{Attack Success Rate.} ASR is defined as the percentage of the attack efforts that make the victim model misclassify the instances that are originally correctly classified. Mathematically, $ASR = \frac{\sum_t{\delta}(\hat{y}^{\prime}_t\neq y_t)}{\sum_t\delta(\hat{y}_t= y_t)}$, where $\hat{y}_t$ is the unperturbed model prediction, $\hat{y}^{\prime}_t$ the model prediction with perturbation, and $y_t$ the ground-truth label. ASR characterizes the capability of the attack model, and higher the ASR, the better the attack. 
\begin{table*}[ht]
\centering
% \begin{tabular}{lC{1.3cm}C{1.3cm}C{1.3cm}C{1.3cm}|C{1.3cm}C{1.3cm}C{1.3cm}C{1.3cm}}
\begin{tabular}{lC{1.2cm}C{1.2cm}C{1.2cm}C{1.2cm}|C{1.2cm}C{1.2cm}C{1.2cm}C{1.2cm}}
\toprule
\multicolumn{1}{c}{\multirow{2}{*}{\textbf{Model}}} & \multicolumn{4}{c|}{\textbf{ASR(\%)}} & \multicolumn{4}{c}{\textbf{F1}} \\ \cmidrule(l){2-9} 
\multicolumn{1}{c}{}                                & NA   & RA   & JO   & AGO  & NA  & RA  & JO & AGO \\ 
\hline
Stocknet & 0 & 3.6  & 12.1 & 11.0 & 1 & 0.97 & 0.89 & 0.89 \\
FinGRU   & 0 & 4.0  & 10.2 & 10.6 & 1 & 0.96 & 0.85 & 0.91 \\
FinLSTM  & 0 & 11.9 & 12.1 & 11.6 & 1 & 0.89 & 0.89 & 0.89 \\
% HAN      & 0 & 3.3  & 3.2  & 3.9     & 1 & 0.97 & 0.97 &      \\ 
\bottomrule
\end{tabular}
\caption{{Results for concatenation attack with deletion perturbation and budgets 1.} NA and RA stand for no attack and random attack respectively, serving as benchmarks. }
\label{tab:appendx_deletion_results}
\end{table*}

\begin{figure*}[ht]
    \centering
    \includegraphics[width=0.9\textwidth]{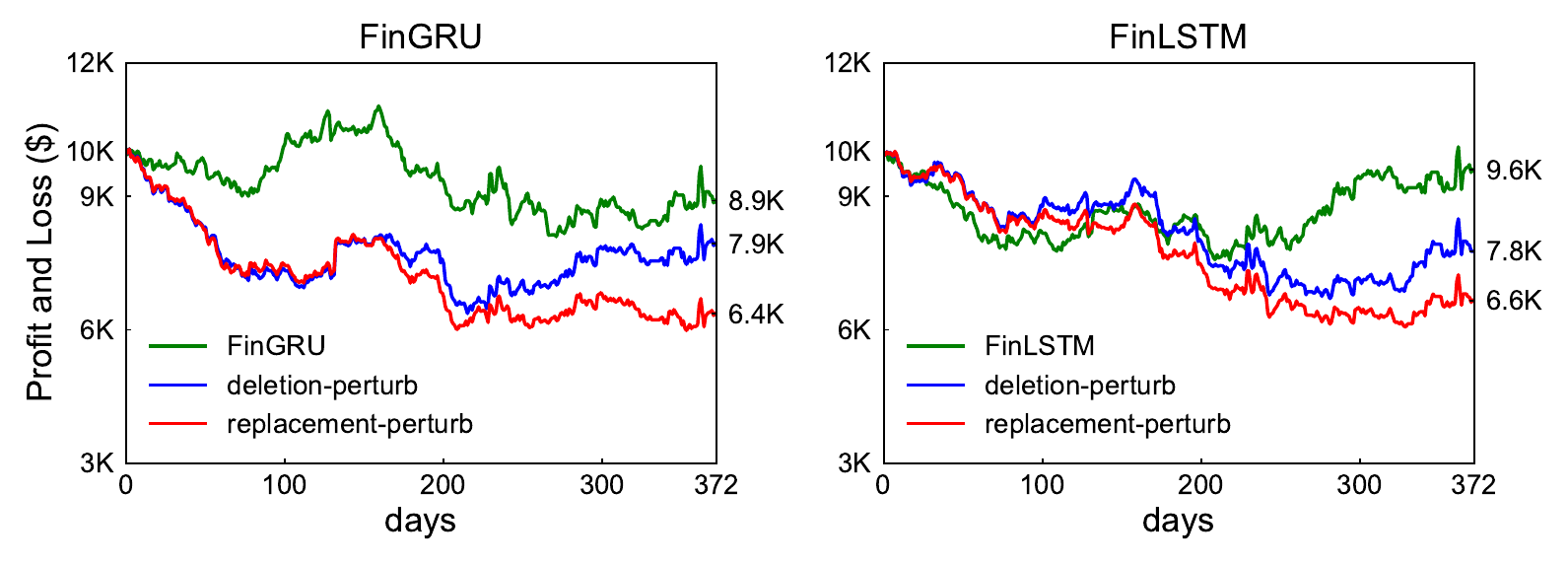}
    \caption{Effect on \textit{ Profit and Loss }of various perturbation methods on FinGRU and FinLSTM. }
    \label{fig:pnl_deletion}
\end{figure*}

\paragraph{F1 Score.} F1 gauges the prediction performance of the victim models. Since we only consider the samples that are correctly predicted, the F1 score in the case of no attack is 1. Apparently, the drop of the F1 score of caused by the perturbation demonstrates the performance of the attack method. 
Unlike ASR, the drops of F1 score gauge the direct impact on the model performance: more successful attack leads to lower post-attack F1 score. 

\paragraph{Profit and Loss.} This widely-used financial indicator measures the profitability of a trading strategy. Assume that the initial net values are \$10K (100\%), accumulate profit and loss for each trade, we can then calculate the final net value of the portfolio and \textit{profit and loss}. A binary financial forecast model can be exploited in many ways, and support various trading strategies, which usually lead to different PnLs. In this paper, we use a simple \textbf{\textit{Long-Only Buy-Hold-Sell}} strategy \cite{sawhney2021fast, feng2019temporal}. More specifically, we  \textit{buy} stock(s) on Day $T$ if the model predicts these stocks go up on Day $T+1$, \textit{hold} for one day, and \textit{sell} these stocks the next day no matter what prices will be, and repeat it. We do not \textit{short} a stock even if the model predicts a negative move in the second day. 

Besides, when the model makes positive prediction on more than one stocks, the money is evenly invested to the stock pool of positive prediction. For example, suppose that we stand on day 4 with portfolio value \$12K. If the model gives positive prediction on 10 of 88 stocks for day 5, we invest 10\% of the total wealth (\$1.2K) to each stock, and sell them at closing prices of day 5. The process continues until the end of the test periods, and the resulting net value of the portfolio is used to calculate the profit and loss of the underlying model. 

The buy-hold-sell strategy monetizes the prediction performance of financial forecast models by betting on the their predictions. The PnL reflects the profitability of the underlying models, even if it is usually influenced by many other confounding factors. Most importantly,  the changes of PnLs caused by perturbation on the victim models only gauge the monetary consequence of our attack, since all else are equal.  

\section{Supplemental Experiment Results}
\label{sec:appendix}
\subsection{Replacement vs deletion perturbation. }
\begin{figure*}[ht]
    \centering
    \includegraphics[width=0.9\textwidth]{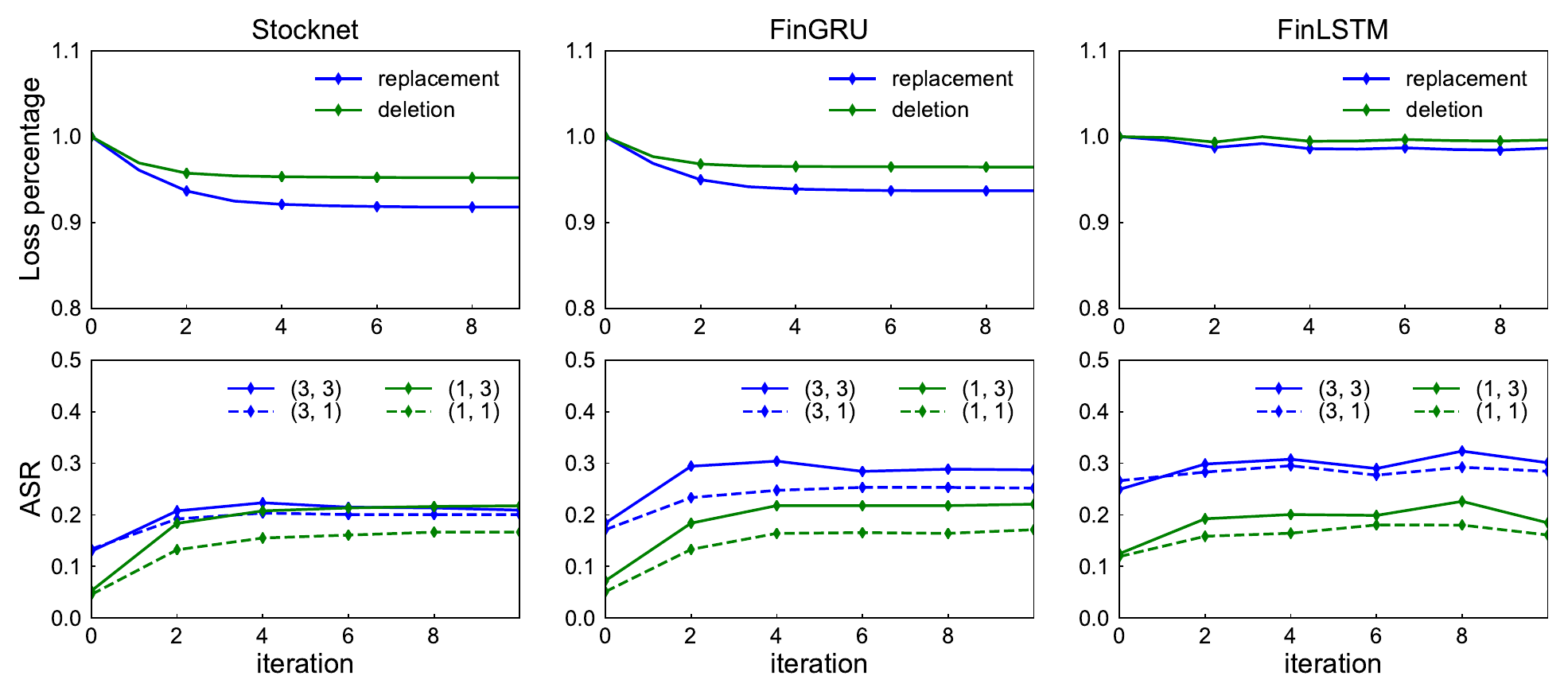}
    \caption{Iteration number effect on prediction loss and attack success rate. The three plots on the first row show the loss trajectory during optimization for the three victim models, and the bottom row reports the ASRs trajectory. The legends for the bottom-row charts read as (tweet budget, word budget).}
    \label{fig:iteration_effect}
\end{figure*}

We report results for concatenation attack with only the \textit{replacement perturbation} in the main text in Table \ref{tab:overall_results}. Here we also report results for the \textit{deletion perturbation} in Table \ref{tab:appendx_deletion_results}. Attacks conducted via deletion perturbation in general perform worse than the results of replacement perturbation. 
We observe ASRs via JO and AGO fall by 5.1\% and 4.1\% respectively compared with the replacement perturbation. 
Accordingly, F1 slightly increases as attack performance worsens. 
There is no significant difference between the two optimizers (JO and AGO) in the case of deletion perturbation, but JO is preferable in terms of optimization efficiency. 

Moreover, we also simulate the trading profit and loss based on FinGRU and FinLSTM. For the sake of consistency, the two models are under concatenation attack with replacement perturbation. Same as our main results, the attack is optimized by JO solver. The simulation results are reported in Figure \ref{fig:pnl_deletion}, which provides further evidence for the potential monetary loss caused by our adversarial attack. Replacement perturbation again outperforms deletion perturbation in the case of FinGRU and FinLSTM.
% It is surprising that the effect of attack on PnL is much more moderate under the deletion perturbation than replacement perturbation, even if deletion attack still achieve fair good attack performance. In fact, the attack basically has no impact on PnL in the case of FinGRU and FinLSTM, even if the attack achieves ASR of 10.2\% and 12.1\% on the two models.  The results suggest that attack performance doesn't always transfer into PnL loss in a linear way. Adversaries should be focus more on the ultimate goal if they target at trading loss. 

\subsection{Effect of Iteration Number}

We experiment with the optimizer to perform gradient descent or greedy search for up to 10 rounds before yielding the final solution. To visualize the effect of iteration, we plot the loss trajectory and ASR along with the optimization iterations in Figure \ref{fig:iteration_effect}. We also collect the average model loss of attack instances at each iteration, and then normalize the loss to set the initial loss as 1. Therefore, the loss trajectory visualization reveals the percentage loss drop during the optimization. We consider two different perturbations (replacement and deletion) under concatenation attacks. The attack is optimized with the JO solver. 

The three charts on the first row of Figure \ref{fig:iteration_effect} show that optimizations on all three victim models quickly converge after 4 iterations in our experiment. Accordingly, ASRs rise gradually during the first 4 iterations, but then flattens or even slides afterward. Such results suggest that our solvers can find the convergence in just a few iterations. Therefore, it makes our attack computationally effective, and insensitive to hyperparameter of iteration number.  

\section{Regularization on Attack Loss.}

The experiment results reported in the main text are generated with the sparsity regularization. We also run ablation experiments that remove sparsity regularization. The results are consistent with our conclusion.
Furthermore, inspired by \cite{srikant2021generating}, we try smoothing attack loss to stabilize the optimization. 
We add Gaussian noise to optimization variables and evaluate the attack 10 times. 
The loss average is then used as the final loss for back-propagation.  
The results show that loss smoothing does not contribute to attack performance in our experiment as it does in \citet{srikant2021generating}.

\section{Attack Word Analysis}
\begin{figure}[ht]
    \centering
    \includegraphics[width=0.45\textwidth]{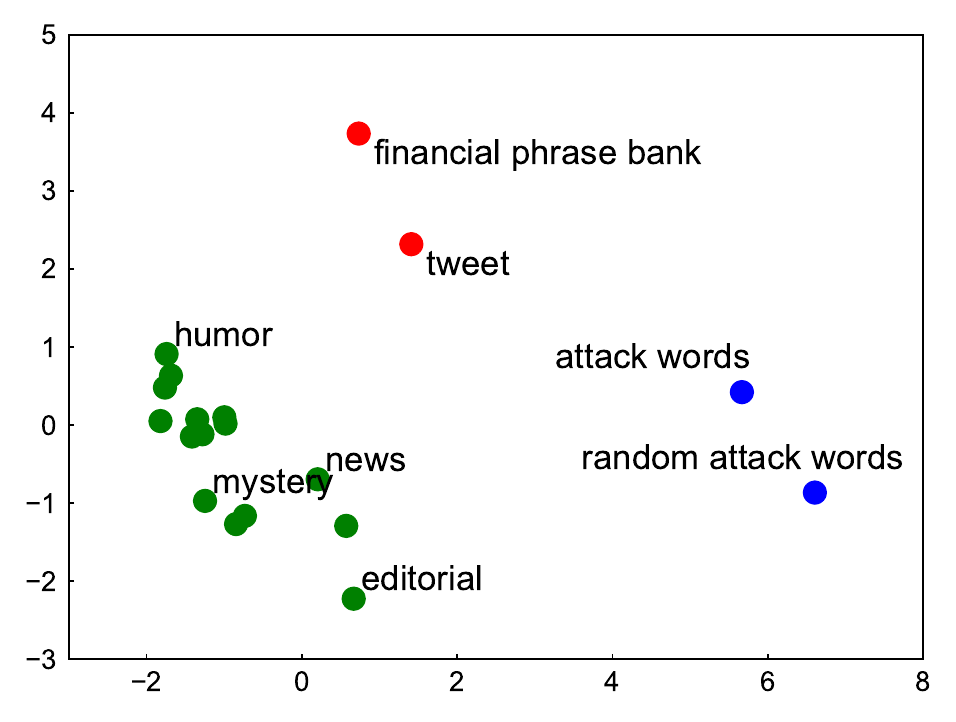}
    \vspace{-10pt}
    \caption{Corpora clusters. 18 corpora are grouped into 3 clusters based on features from LIWC. In order to visualize the clusters, principal component analysis is applied to the features to find the first 2 principal components, which are then used as x-axis and y-axis to generate this figure. }
    \vspace{-10pt}
    \label{fig:word_cluster}
\end{figure}

To qualitatively understand what kinds of words and tweets are being selected in the perturbation and retweet, 
% we run our experiment in the domain of finance, and it is interesting to know whether financial adversarial attack exhibits any patterns specific to the domain. 
we compare our tweet corpus and the selected word replacements with 15 corpora of different genres in Brown corpus via Linguistic Inquiry and Word Count program (LIWC) \cite{tausczik2010psychological}. As Brown corpus does not have a financial genre, we also use Financial Phrase Bank \cite{Malo2014GoodDO}. 
% We find that the sample financial corpora load significantly in the categories of \textit{analytic}, \textit{positive}, \textit{certain}, \textit{word}, \textit{reward}, \textit{drive}, \textit{achieve} and \textit{money} among others. 
% Those categories reflect the characteristics of financial texts: more analytic in nature, focusing on monetary side and often polarized. On the other hand, the attack words is close to financial corpora, but are even more polarized (both positive and negative) and \textit{present} and \textit{future} focused. 
We then run K-means clustering on these 18 corpora based on the feature matrix from LIWC. 
As shown in Figure \ref{fig:word_cluster}, financial corpora (red), Brown general word corpus (green), and attack words (blue) are grouped into three clusters, indicating the inherent difference of those text genres. 
Moreover, we observe that target words identified by our solvers (red ``tweet'' and blue ``attack words'' dots) are closer to financial corpora than ``random attack words''. 
% As a result, our attack is more likely to evade attack detection mechanism than random attack. Furthermore, perturbation on polarized words to either increase or decrease polarization of articles is more likely to alter human perception and decision, our results, however, imply that perturbing polarized words may not make successful attack even if it is intriguing to do so. 

\section{Example of Adversarial Quote Tweet}

Table \ref{tab:adver_example} reports 10 adversarial quote tweets  generated in concatenation attack mode with JO and AGO solver and replacement perturbation. For all the examples, the victim model predicts positive outcomes originally, and but predicts negative outcomes after adding the adversarial quote tweet. 

\begin{table*}[ht]
    \begin{tabular}{p{0.9\linewidth}}
    \hline 
    \hline
    {\textbf{Adversarial Quote Tweets Generated by AGO}} \\
    \hline
    \textbf{Benign tweet:} \$BHP announces the demerger of its non-core assets - details expected to be \textcolor{green}{filled} in on Tuesday. \\
    \textbf{Adversarial quote tweet:} \$BHP announces the demerger of its non-core assets - details expected to be \textcolor{red}{exercised} in on Tuesday \\
    \hline
    
    \textbf{Benign tweet:} \textcolor{green}{Mover} and Shakers... Losers- \$KO \$ABX \$DD. Winners- \$LAND \$CHL \$BHP.\\
     \textbf{Adversarial quote tweet:} \textcolor{red}{Shoulder} and Shakers... Losers- \$KO \$ABX \$DD. Winners- \$LAND \$CHL \$BHP.\\ 
     \hline
     \textbf{Benign tweet:} Latest \textcolor{green}{information} on \#stocks like \$TDS \$DIS \$CPWR \$BLOX Give it a try. 
     \textbf{Adversarial quote tweet:} Latest \textcolor{red}{advance} on \#stocks like \$TDS \$DIS \$CPWR \$BLOX Give it a try. \\
     \hline
     \textbf{Benign tweet:} Top \textcolor{green}{Buy} Stocks for July 29: \$AMX, \$EXC, \$NLY, \$ORCL, \$XRX.\\
      \textbf{Adversarial quote tweet:}  Top \textcolor{red}{unsettled} Stocks for July 29: \$AMX, \$EXC, \$NLY, \$ORCL, \$XRX.\\
      \hline
      \textbf{Benign tweet:} \$JKHY \#Stocks you might  \textcolor{green}{want} to buy \$HRS \$SRCL \$PCLN \#singedinvestor.
      \textbf{Adversarial quote tweet:} \$JKHY \#Stocks you might \textcolor{red}{confirm} to buy \$HRS \$SRCL \$PCLN \#singedinvestor.  \\
    \hline
    \hline
    {\textbf{Adversarial Quote Tweets Generated by JO}} \\
    \hline
    \textbf{Benign tweet:} \#Stocks you migh \textcolor{green}{want} to Sell \$CERN \$CAT \$PX \$DO Try this {http:\slash \slash t.co\slash 95PUim108L}. \\
    \textbf{Adversarial quote tweet:} \#Stocks you migh \textcolor{red}{announce} to Sell \$CERN \$CAT \$PX \$DO Try this \text{http:\slash \slash t.co\slash 95PUim108L}.\\
    \hline
    \textbf{Benign tweet:} The Most Accurate Stock \textcolor{green}{Alert} Service Made \$240,000 Yesterday http:\slash \slash t.co\slash 8WvnrVmkBS .. \$SNY \$SO \$SOL. \\
    \textbf{Adversarial quote tweet:} The Most Accurate Stock \textcolor{red}{preserving} Service Made \$240,000 Yesterday http:\slash \slash t.co\slash 8WvnrVmkBS .. \$SNY \$SO \$SOL.
    \\
    \hline
    \textbf{Benign tweet:} Is it \textcolor{green}{breakout} time on \$HON \$CF \$LEG \$IBM Give it a try.  \\
    \textbf{Adversarial quote tweet:} Is it \textcolor{red}{appealing} time on \$HON \$CF \$LEG \$IBM Give it a try.  \\
    \hline
    \textbf{Benign tweet:} Latest \textcolor{green}{information} on \#stocks like \$PETM \$UTX \$BRCD \$CI \#moneymanagement. \\
    \textbf{Adversarial quote tweet:}  Latest \textcolor{red}{discovery} on \#stocks like \$PETM \$UTX \$BRCD \$CI \#moneymanagement. \\
    \hline
    \textbf{Benign tweet:} \$BABA actually showing signs of life...would love a move back toward 90 although seems unlikely at \textcolor{green}{moment}.\\
    \textbf{Adversarial quote tweet:} \$BABA actually showing signs of life...would love a move back toward 90 although seems unlikely at \textcolor{red}{playday}.\\
    \hline
    \hline
    \end{tabular}
    \caption{Ten examples of adversarial quote tweets generated by the concatenation attack}
    \label{tab:adver_example}
\end{table*}

\end{document}